\documentclass[superscriptaddress,nobibnotes,amsmath,amssymb,notitlepage,twocolumn,pra,longbibliography]{revtex4-2}

\usepackage{bm,braket}
\usepackage[toc,page]{appendix}
\usepackage{comment}
\usepackage{dcolumn}
\usepackage{transparent}

\usepackage{amsfonts}

\usepackage{graphicx,color,hyperref}
\usepackage[caption=false]{subfig}
\hypersetup{colorlinks=true, linkcolor=blue, citecolor=blue, urlcolor=blue} 

\graphicspath{{./img/}}

\newcommand{\beq}[1]{\begin{equation}\label{#1}}
\newcommand{\eep}{\;.\end{equation}}
\newcommand{\eec}{\;,\end{equation}}
\newcommand{\eeq}{\end{equation}}

\DeclareMathAlphabet{\mathcal}{OMS}{cmsy}{m}{n} 

\newcommand{\hn}{\hat{\boldsymbol{n}}}
\newcommand{\hmm}{\hat{\boldsymbol{m}}}

\usepackage{amsmath}
\usepackage{amssymb}
\usepackage{xcolor}
\usepackage{bbm}
\usepackage{physics}
\usepackage{float}
\usepackage{graphicx}
\usepackage{dcolumn} 
\usepackage{bm} 
\usepackage{siunitx}
\usepackage{enumitem}  
\usepackage{tikz}
\usetikzlibrary{quantikz2}

\makeatletter
\renewcommand*{\fnum@figure}{{\normalfont\bfseries \figurename~\thefigure}}
\makeatother

\allowdisplaybreaks

\definecolor{orange}{rgb}{1,0.5,0}

\graphicspath{{./img/}}

\usepackage{tikz}
\usetikzlibrary{quantikz2}

\DeclareMathAlphabet{\mathcal}{OMS}{cmsy}{m}{n} 

\makeatletter
\newcommand{\specificthanks}[1]{\@fnsymbol{#1}}
\makeatother

\begin{document}

\preprint{APS/123-QED}

\title{Estimating the Fisher information from ARPES in general two-band models}

\author{Gunnar F. Lange}
\email{gunnalan@uio.no}
\affiliation{Department of Physics, University of Oslo, N-0316 Oslo, Norway}
\affiliation{Centre for Materials Science and Nanotechnology, University of Oslo, N-0316 Oslo, Norway}

\date{\today}

\begin{abstract}
The Quantum Fisher information is frequently studied in the fields of quantum sensing and quantum metrology, as it specifies the sensitivity of a sensor through the Crámer-Rao bound, and in condensed matter physics where it relates to the quantum metric. Measuring the quantum Fisher information is therefore of significant current interest. In this work, we propose a scheme for measuring the quantum Fisher information with respect to any parameter in a general two-band solid state system using angle-resolved photoemission spectroscopy with carefully tuned incidence angles and light polarization. We investigate conditions under which this can be related back to information about the ground state of the material itself changing with system parameters.
\end{abstract}
\maketitle

\section{Introduction}
A key challenge in quantum materials and quantum technologies is characterizing how a quantum state varies with the parameters of the system. This question unifies many areas of quantum science, including quantum sensing and metrology \cite{wiseman_quantum_2009, degen_quantum_2017}, where state variation defines sensing capability, and condensed matter physics, where state variations lead to both geometric and topological effects \cite{facchi_classical_2010, bouhon_geometric_2020,torma_essay_2023,yang_berry_2026}.

One particular instance of this, which has received significant attention in recent years, is the quantum metric of a solid-state system \cite{torma_essay_2023}, which captures variations of the ground state with respect to Bloch momentum $\boldsymbol{k}$. Various proposals have recently been put forward to measure and simulate the quantum metric  \cite{laurell_witnessing_2025}, including in inelastic X-ray scattering \cite{balut_quantum_2025,balut_fundamental_2026}, scanning tunneling microscopy \cite{zhang_spin-polarized_2026}, NV-centers \cite{yu_experimental_2024} and in angle-resolved photoemission spectroscopy (ARPES) \cite{kim_direct_2025}. ARPES is a particularly appealing technique to measure the quantum metric, as it gives access to $\boldsymbol{k}$-resolved maps of the quantum metric, which play a role in optical properties \cite{ahn_riemannian_2022}, superfluid weight coupling \cite{huhtinen_revisiting_2022} as well as local bounds relating geometry to topology \cite{onishi_fundamental_2024, onishi_quantum_2025,jankowski_quantum_2025, telle_optimally_2026}. Previous work using ARPES has focused primarily on measuring the metric in a small region of $\boldsymbol{k}$-space for two-band systems with sublattice symmetry.

The quantum metric is, however, only one special case of the parameter dependence of a quantum state. The general sensitivity of a quantum state to a parameter is precisely captured by the (quantum) Fisher information \cite{braunstein_statistical_1994}. The classical Fisher information (CFI) originates in traditional probability theory, where it measures the variation of a probability distribution $p(x;\lambda)$, dependent on some parameter $\lambda$. The extension to the quantum Fisher information (QFI) arises when taking the supremum of the CFI over all possible measurement operators \cite{wiseman_quantum_2009}. The CFI/QFI view on parameter dependence and the quanutm metric therefore unifies solid state and quantum sensing interests. In quantum sensing and parameter estimation, sensor performance is commonly measured against the QFI with respect to some external parameter (strain, magnetic field etc.), as this ensures high sensitivity through the Crámer-Rao bound \cite{cramer_mathematical_1999, radhakrishna_rao_information_1945}.

In this work, we employ the connection between ARPES and Pauli measurements in two-band systems, as well as the connection between the CFI and the QFI, to systematically explore how to measure Fisher information in an ARPES experiment. This generalizes previous results in two ways: first of all, it views measuring the quantum metric as only a special case of a more general problem of measuring parameter dependence of the quantum state, opening the door to estimating the Fisher information with respect to other properties (such as strain) in a solid-state material, which is of interest in quantum sensing \cite{zalandauskas_theory_2026}. Secondly, because our method does not rely on state tomography, it can be extended beyond systems with sublattice symmetry, enabling the estimation of both the quantum metric and more general quantum Fisher information measures in general two-band models.

A general issue with using ARPES to estimate parameter dependence, is that the ARPES matrix elements themselves can become parameter dependent, particularly when scattering is taken into account. We explore this quantitatively, providing concrete predictions of where in $\boldsymbol{k}$-space we can expect to access the QFI from an ARPES measurement using our technique.

\begin{figure}[t!]
    \centering
    \def\svgwidth{0.95\linewidth}
    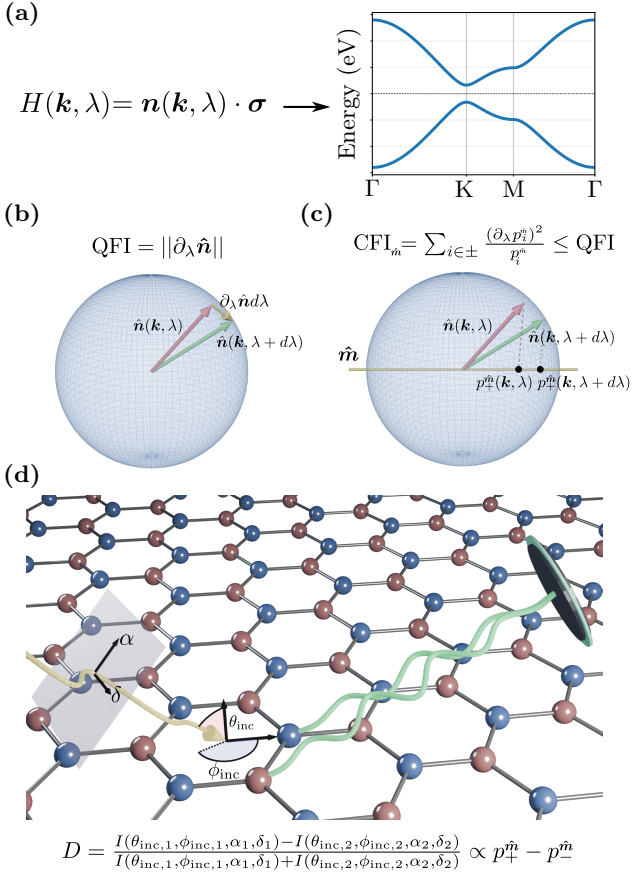
    \caption{Conceptual summary of the work. \textbf{(a)} We study general two-band Bloch Hamiltonians $H(\boldsymbol{k},\lambda)$, depending on a parameter $\lambda$. \textbf{(b)} The sensitivity of the ground state with respect to  $\lambda$ is captured by the Quantum Fisher Information (QFI). \textbf{(c)} The QFI can be approximated by the Classical Fisher Information (CFI) along some axis $\hmm$, (CFI$_{\hmm}$), corresponding to a Pauli measurement along $\hmm$. \textbf{(d)} These probabilities can be accessed by Angle-Resolved Photoemission Spectroscopy (ARPES) by suitably tuning incidence angles $(\theta_{\mathrm{inc}}, \phi_{\mathrm{inc}})$ and light polarizations $(\alpha,\delta)$. For certain axis choices $\hmm_{\mathrm{opt}}$ and $\boldsymbol{k}$-values, CFI$_{\hmm_{\mathrm{opt}}}$ equals the QFI, and the QFI can be reconstructed from a dichroism coefficient $D$.}
    \label{fig:conceptual_fig}
\end{figure}

We illustrate this framework in two-dimensional graphene and hexagonal boron phosphide monolayers, calculating both the quantum metric and the strain response. The rest of this work is structured as follows: In Sec.~\ref{sec:background_theory}, we recap the basic theory required and illustrate how ARPES intensities can be mapped to the CFI. In Sec.~\ref{sec:methods} we discuss how to get to the QFI, and also details of what materials we consider. In Sec.~\ref{sec:results} we show results for the QFI for both momentum and strain responses in two materials, both with and without changing the orbital matrix elements. These results are discussed in Sec.~\ref{sec:discussion}, before we conclude in Sec.~\ref{sec:Conclusion}. The main ideas of this work are summarized in Fig.~\ref{fig:conceptual_fig}.

\section{Theory}\label{sec:background_theory}
\subsection{Background: Fisher information for two-band systems}\label{sec:QFI_CFI_Theory}
Consider a two-band system described by a Bloch Hamiltonian $H(\boldsymbol{k},\lambda) = \boldsymbol{n}(\boldsymbol{k},\lambda)\cdot \boldsymbol{\sigma}$, with $\boldsymbol{\sigma}=\{\sigma_x,\sigma_y, \sigma_z\}$, $\boldsymbol{k}$ the Bloch momenta and $\lambda$ any additional parameter on which the system depends. In our context, $\lambda$ will be either a specific Bloch momentum component or strain. At low temperature, the ground state can be written as:
\begin{equation}
    P(\boldsymbol{k},\lambda) = \ket{u}\bra{u} = \frac{1}{2}(1-\hat{\boldsymbol{n}}(\boldsymbol{k},\lambda)\cdot \hat{\boldsymbol{\sigma}}),
\end{equation}
with $\hat{\boldsymbol{n}} \equiv \boldsymbol{n}/||\boldsymbol{n||}$. The (diagonal part of the) quantum metric with respect to $\lambda$ is given by \cite{mitscherling_gauge-invariant_2025}:
\begin{equation}
    g_{\lambda\lambda} = \frac{1}{2}\mathrm{Tr}[\partial_{\lambda}P\partial_{\lambda}P] = \frac{1}{4}||\partial_{\lambda}\boldsymbol{\hat{n}}||^2.
\end{equation}
Using the well-known \cite{facchi_classical_2010} equivalence between the quantum metric and the QFI for pure states, $\mathrm{QFI} = 4g$ gives:
\begin{equation}\label{eq:QFI_def_two_band}
    \mathrm{QFI} = ||\partial_{\lambda}\hat{\boldsymbol{n}}||^2.
\end{equation}
The QFI characterizes the maximum information about $\lambda$ that can be extracted from $P$, and is given by \cite{wiseman_quantum_2009}:
\begin{equation}\label{eq:QFI_def_CFI}
    \mathrm{QFI} = \mathrm{sup}_{\{\Pi_i\}} \mathrm{CFI}_{\{\Pi_i\}} = \mathrm{sup}_{\{\Pi_i\}}\sum_{i}\frac{(\partial p_{\lambda}^{\Pi_i})^2}{p_{\lambda}^{\Pi_i}},
\end{equation}
where $\{\Pi_i\}$ is a set of positive operator-valued measures (POVMs) and $p_{\lambda}^{\Pi_i} = \mathrm{Tr}[\Pi_iP(\lambda)]$ at every $\boldsymbol{k}$. Thus, the QFI is the supremum over all POVMs of the classical Fisher information (CFI). A set of POVMs that satisfy $\mathrm{CFI}_{\{\Pi_i\}}=\mathrm{QFI}$ form an \textit{optimal} measurement. The optimal POVMs are not generically unique, but one optimal basis is always given by the eigenstates of the symmetric logarithmic derivative (SLD), which in the present case takes the form $\mathrm{SLD} = \partial\hat{\boldsymbol{n}}\cdot \boldsymbol{\sigma}$.

To connect with photoemission theory, we focus on the simplest case of Pauli measurements along a fixed axis $\hmm$, i.e. we consider measurement operators of the form:
\begin{equation}\label{eq:definition_POVM_Pauli}
\Pi_{\pm} = \frac{1}{2}(1\pm \hmm\cdot\boldsymbol{\sigma}),
\end{equation}
with associated probabilities:
\begin{equation}
p_{\pm}(\lambda) = \mathrm{Tr}[\Pi_{\pm}P(\lambda)] = \frac{1\mp \hn\cdot\hmm}{2}.
\end{equation}
The associated CFI$_{\hmm}$ for a measurement along $\hmm$ is then:
\begin{equation}
\mathrm{CFI}_{\hmm}=\frac{([\partial\hn\cdot \hmm]^2 + 2[\partial\hn\cdot\hmm][\hn\cdot \partial \hmm]+[\hn\cdot \partial \hmm]^2)}{1-(\hn\cdot\hmm)^2}.
\end{equation}
To investigate only the information contained in the ground state itself, the measurement axis should carry no information about the parameter $\lambda$. Therefore, restricting to $\partial_{\lambda} \hmm = 0$, the CFI simplifies to:
\begin{equation}\label{eq:CFI_general_axis}
\mathrm{CFI}_{\hmm}=\frac{(\partial\hn\cdot \hmm)^2 }{1-(\hn\cdot\hmm)^2}.
\end{equation}

It follows that we can interpret the QFI as a Pauli measurement, and by picking e.g. locally $\hmm \parallel \partial \hn$, we can ensure that CFI$_{\hmm} = $ QFI. This optimal axis is not unique as we discuss further in App.~\ref{ap:optimal_axis_general}.

Although this is a promising avenue to measure the QFI, it is important to note that in Eq.~\eqref{eq:definition_POVM_Pauli} we considered a local measurement in $\boldsymbol{k}$ Pauli in the \textit{orbital space} of the two-band model. Such a measurement is generally very difficult to perform. We now discuss how to perform such a measurement using angle-resolved photoemission spectroscopy (ARPES), by carefully varying polarization and incidence angles.

\subsection{Photoemission matrix elements for two-band systems}
ARPES is well-suited for this problem, as it is the main technique to accessing $\boldsymbol{k}$-resolved properties such as band structures. Furthermore, in recent years, there has been significant interest in orbital- or spin-resolved ARPES, where a judicious choice of polarization and incidence angle allows access to specific properties of the quantum state \cite{beaulieu_revealing_2020,boban_scattering_2025,hwang_direct_2011,moser_toy_2023,plucinski_origin_2023,schuler_polarization-modulated_2022, gierz_illuminating_2011}. We first discuss the general form of the ARPES response, and then explore how ARPES can be used to perform effective Pauli measurements, accessing CFI$_{\hmm}$. 

For concreteness, assume that $H(\boldsymbol{k},\lambda)$ is a tight-binding model with two orbitals per primitive unit cell, labelled by $n\in \{A,B\}$, under the standard tight-binding approximation of point-like orbitals which we assume to be localized on the atoms at position  $\boldsymbol{\tau}_n$. We define the home-cell basis functions $\phi_{n,\boldsymbol{0}}(\boldsymbol{r})$, as well as their periodic images $\phi_{n,\boldsymbol{R}} = \phi_{n,\boldsymbol{0}}(\boldsymbol{r}-\boldsymbol{R}-\boldsymbol{\tau}_{n})$. The Pauli measurement corresponds to measuring the ($\boldsymbol{k}$-resolved) relative occupation and phase between these two orbitals. We define the Fourier-transformed basis orbitals as:
\begin{equation}\label{eq:basis_def}
|n,\boldsymbol{k}\rangle =\frac{1}{\sqrt{N}}\sum_{\boldsymbol{R}}e^{i\boldsymbol{k}\cdot(\boldsymbol{R}+\boldsymbol{\tau}_{n})}|\phi_{n},\boldsymbol{R}\rangle,
\end{equation}
i.e. we choose the convention in which our Hamiltonian explicitly includes orbital embeddings \cite{vanderbilt_berry_2018}. The Bloch eigenstates of the occupied ($-$) and unoccupied ($+$) bands are then:
\begin{equation}
    |\psi_{\boldsymbol{k}}^{\pm}\rangle=\sum_{n\in \{A,B\}} C_n^{\pm}|n, \boldsymbol{k}\rangle,
\end{equation}
where $C_n^{\pm}(\boldsymbol{k})$ are the state coefficients, obtained from diagonalizing the Bloch Hamiltonian. We now consider a photoemission experiment where a photon with energy $\hbar \omega$ knocks out an electron through a photoinozation process, as sketched in Fig.~\ref{fig:conceptual_fig}(d). Following \cite{moser_experimentalists_2017}, we assume rapid photoionization, so that the emitted photoelectron decouples from the electrons in the material. The one-electron transition matrix element to a final state $|\psi_f\rangle$ can then be written as:
\begin{equation}\label{eq:matrix_elements_velocity_gauge}
    M(\boldsymbol{k}_f) = \langle \psi_f|\boldsymbol{A}\cdot \boldsymbol{p}|\psi_{\boldsymbol{k}}^-\rangle.
\end{equation}

 We will approximate the final state as a scattered wave, originating from an emitter site in a unit cell, as is implemented in the Electron Difraction in Atomic Clusters (\texttt{EDAC}) code \cite{garcia_de_abajo_multiple_2001}. Scattering is important \cite{boban_scattering_2025}, as it e.g. allows orbitals with zero inital-state orbital angular momentum (OAM) to become dependnet on circular dichroism at normal incidence through the Diamon effect \cite{daimon_strong_1993}. Thus, to fully understand where orbital dependence of ARPES spectra originate, scattering should be accounted for. Following \cite{boban_scattering_2025}, Eq.~\eqref{eq:matrix_elements_velocity_gauge} can be approximated as:
\begin{equation}\label{eq:matrix_element_full_expression}
    M(\boldsymbol{k}_f) = \sum_n C_n^-(\boldsymbol{k}_{\parallel})M_n(\boldsymbol{k}_{f}),
\end{equation}
with $\boldsymbol{k}_f$ the wavector associated with the outgoing electron, and $\boldsymbol{k}_{\parallel}$ the associated in-plane component. This assumes in-plane momentum conservation, and we relabel $\boldsymbol{k}_{\parallel}$ to $\boldsymbol{k}$ going forward. The orbital matrix elements are given by:
\begin{equation}\label{eq:atomic_matrix_elements_spherical}
M_n(\boldsymbol{k}_f) = \langle \psi_f|\boldsymbol{e}\cdot\boldsymbol{r}|\phi_n,\boldsymbol{0}\rangle = \sum_{\mu\in \{-1,0,1\}}\boldsymbol{e}_{\mu}\cdot \langle \psi_f|\boldsymbol{r}|\phi_{n,\boldsymbol{0}}\rangle,
\end{equation}
where $\boldsymbol{e}_{\mu}$ is the polarization vector of light decomposed into spherical harmonics, $\boldsymbol{r}$ is the position operator and we work in length gauge as we assume the basis orbitals to be well localized. We can define the matrix element components $\boldsymbol{M}_n = [M_n^{-1}, M_n^{0}, M_n^{+1}]^T$, so that for any polarization vector $\boldsymbol{e}$, the total matrix element is:
\begin{equation}
 M_n = \boldsymbol{e}^T\boldsymbol{M}_n   
\end{equation}

From Fermi's golden rule, the photoemission intensity probed by ARPES is proportional to \cite{moser_experimentalists_2017,boban_scattering_2025}
\begin{equation}\label{eq:def_ARPES_intensity}
I(E, \boldsymbol{k}_f, \omega) \propto |M(\boldsymbol{k}_f,\omega)|^2\delta(\hbar \omega-\epsilon_{\boldsymbol{k}}-W-E_B),
\end{equation}
where $\epsilon_{\boldsymbol{k}}$ is the occupied band energy at momentum $\boldsymbol{k}$, $\omega$ is the frequency of the incoming photon, $W$ is the work function and $E_B$ the binding energy. In writing this proportionality, we have ignored the one-electron removal spectral function \cite{moser_experimentalists_2017}, which encodes many-body effects. To understand which information about the ground state is \textit{in principle} accessible by ARPES, we focus on the matrix element $M(\boldsymbol{k}_f, \omega)$, which encode the single-particle state dependence, and ignore the spectral functions and $\delta$-function (as well as thermal broadening and other noise) in what follows. We further focus on a single photon energy $\hbar \omega =40$\,eV.

\subsection{Photoemission processes as Pauli measurements}\label{sec:photoemission_as_Pauli}
To link the ARPES matrix element $M(\boldsymbol{k})$ to a Pauli measurement, and ultimately to the QFI, we define the \textit{effects} from the matrix elements in Eq.~\eqref{eq:atomic_matrix_elements_spherical} as:
\begin{equation}\label{eq:effects}
\boldsymbol{v}(\boldsymbol{k}) = \begin{pmatrix}
    M_A(\boldsymbol{k})\\
    M_B(\boldsymbol{k}) 
\end{pmatrix} = \begin{pmatrix}
    \boldsymbol{e}^T\boldsymbol{M}_A\\
    \boldsymbol{e}^T\boldsymbol{M}_B
\end{pmatrix}.
\end{equation}
Including the ground state projector $P(\boldsymbol{k}) = \frac{1}{2}(1-\hn(\boldsymbol{k})\cdot\boldsymbol{\sigma})$, the total intensity is proportional to:
\begin{equation}
I(\boldsymbol{k}) \propto |M(\boldsymbol{k})|^2 = \mathrm{Tr}[P\boldsymbol{v}^{\dagger}\boldsymbol{v}].
\end{equation}
We can decompose the effects as $\boldsymbol{v^{\dagger}}\boldsymbol{v} = a\boldsymbol{1} +\boldsymbol{g}\cdot\boldsymbol{\sigma}$ with $a$ proportional to the total intensity:
\begin{equation}\label{eq:def_total_intensity_a}
    a = \frac{1}{2}(|M_A|^2+|M_B|^2)
\end{equation}
And $\boldsymbol{g}$ the measurement axis given by:
\begin{equation}
    \boldsymbol{g} = \begin{pmatrix}\mathrm{Re}(M_AM_B^*)\\ \mathrm{Im}(M_AM_B^*)\\ \frac{1}{2}(|M_A|^2-|M_B|^2)\end{pmatrix},
\end{equation}\label{eq:def_g_in_terms_of_mat}
with $a = ||\boldsymbol{g}|| \geq 0$. Defining $\hat{\boldsymbol{m}} = \boldsymbol{g}/a$ gives:
\begin{equation}
    I(\boldsymbol{k})\propto a(1-\hat{\boldsymbol{n}}\cdot\hat{
    \boldsymbol{m}}).
\end{equation}
Therefore, the matrix element contribution to the ARPES intensity for a two-band system can be related to the projection of the lowest band eigenstate at $\boldsymbol{k}$ onto the axis $\hat{\boldsymbol{m}} = \boldsymbol{g}/a$. At fixed $\boldsymbol{k}$, the axis $\hat{\boldsymbol{n}}$ is fixed by the ground state, whereas the axis $\hat{\boldsymbol{m}}(\boldsymbol{e})$ can be controlled by changing the light polarization $\boldsymbol{e}$. We emphasize that the axis $\hmm$ only depends on the orbital matrix elements $\{\boldsymbol{M}_A, \boldsymbol{M}_B\}$, not on the Bloch eigenvectors and therefore not on the Hamiltonian.

In practice, absolute intensities in ARPES depend on a variety of factors, including sampling time, device specifications and spectral functions, and are therefore not suited to get quantitative estimates of the QFI. Therefore, instead of comparing absolute intensities between different setups to extract orbital information, we consider a general dichroism coefficient, $D$. Defining $I_1\propto a_1(1-\hmm_1\cdot\hn)$ to be the ARPES intensity with polarization vector $\boldsymbol{e}_1$ and similarly for $I_2 \propto a_2(1-\hmm_2\cdot \hn)$ with polarization vector $\boldsymbol{e}_2$, the coefficient $D$ is defined as:
\begin{equation}\label{eq:general_dichroism_def}
    D = \frac{I_1-I_2}{I_1+I_2}.
\end{equation}
When, e.g. $\boldsymbol{e}_1/\boldsymbol{e}_2$ corresponds to left/right circularly polarized light at fixed incidence, then $D$ is the circular dichroism coefficient at that incidence. Focusing again on the matrix elements and rewriting everything in terms of the axis $\hn,\hmm$, Eq.~\eqref{eq:general_dichroism_def} can be written as:
\begin{equation}
    D \propto \frac{a_1-a_2+(a_1\boldsymbol{\hat{m}}_1-a_2\boldsymbol{\hat{m}_2})\cdot\boldsymbol{\hat{n}}}{a_2+a_2+(a_1\boldsymbol{\hat{m}}_1+a_2\boldsymbol{\hat{m}}_2)\cdot \boldsymbol{\hat{n}}}.
\end{equation}
This is not generically equal to a Pauli measurement. In the specific case, however, where $a_1 = a_2$ and $\hat{\boldsymbol{m}}_1 = -\hat{\boldsymbol{m}}_2$, the dichroism coefficient $D = \hat{\boldsymbol{m}}_1\cdot \hat{\boldsymbol{n}} = -\hmm_2\cdot \hn$, so that $D$ can be related to a Pauli measurement. Specifically, we can define
$D = p_+-p_-$ with
\begin{equation}\label{eq:p_plus_p_minus_def}
p_+ =  \frac{1+\hmm_1\cdot \hn}{2}, \quad p_- =  \frac{1-\hmm_1\cdot \hn}{2}.
\end{equation}
Thus, the dichroism coefficient $D$ can be linked to a Pauli measurement in $\boldsymbol{k}$-space of the orbital degree of freedom along $\hmm$, provided that the polarization vectors $\boldsymbol{e}_{1/2}$ are chosen so that the intensities $I_{1,2}$ satisfy: (i) \textit{balanced intensities} ($a_1 = a_2$) and (ii) \textit{opposite axis}, $\boldsymbol{g}_1 = -\boldsymbol{g}_2$.

As shown in Fig.~\ref{fig:conceptual_fig}(d), there are in principle four parameters which can be tuned per polarization vector $\boldsymbol{e}$: the incidence angles $(\theta_{\mathrm{inc}}, \phi_{\mathrm{inc}})$ and the light polarization $(\alpha, \delta)$. This suggests that it may be possible, with sufficient control of polarization and incidence angles, to satisfy the required conditions. We give a constructive algorithm in App.~\ref{ap:finding_optimal_polarizations} for finding such polarizations locally in $\boldsymbol{k}$. The algorithm, which works as long as $\boldsymbol{M}_A \nparallel\boldsymbol{M}_B$, also ensure that the axis $\hmm \propto\partial\hn$ and that $||\boldsymbol{e}_{1,2}||=1$. The only drawback is that it requires exquisite control of polarization and incidence angles, and that it may result in small absolute intensities $a$, potentially necessitating long integration times. Nonetheless, with sufficient control, it is generically possible to find polarizations locally in $\boldsymbol{k}$ which correspond to measurements along the optimal axis. In particular, this is possible for any perturbation $\lambda$, and for an arbitrary two-band model, as long as $M_A(\boldsymbol{k}) \nparallel M_B(\boldsymbol{k})$. This is a key result of our work.

\section{Methods}\label{sec:methods}
\subsection{From Pauli measurement to CFI and QFI}\label{sec:optimal_Pauli}
Having shown that ARPES dichroism elements at a specific $\lambda$ can be made equal to an optimal Pauli measurement for a judicious choice of polarization vectors $\boldsymbol{e}_{1,2}$, we now ask whether the link between QFI and CFI$_{\hmm}$ discussed in Sec.~\ref{sec:QFI_CFI_Theory} can be used to obtain the QFI from these measurements. 

The main issue is that the states $C_n^-(\boldsymbol{k})$ are not the only $\lambda$-dependent quantity in the ARPES intensities. In particular, the orbital matrix elements $M_n(\boldsymbol{k})$ themselves will generically also depend on $\lambda$. Thus, while it is possible to obtain $p_{+}$ and $p_{-}$ from Eq.~\eqref{eq:p_plus_p_minus_def}, their derivatives will contain additional terms which vary with the matrix elements. In the idealized case, where the only variation of $D$ with $\lambda$ is due to changes in the states, the CFI is:
\begin{equation}\label{eq:CFI_to_D}
    \mathrm{CFI}_{\hmm} = \frac{(\partial_{\lambda} D)^2}{1-D^2}.
\end{equation}
In practice, we compute the CFI$_{\hmm}$ from Eq.~\eqref{eq:CFI_to_D} using a finite difference five-point stencil, i.e. evaluating the dichroism coefficient at $\{\lambda, \lambda\pm d\lambda, \lambda \pm 2d\lambda\}$. To ensure that the axis $\hmm$ does not vary with the parameter $\lambda$, $\partial \hmm = 0$, we use the same polarization axis at all $\lambda$'s in the stencil. This then allows us to separate two contributions: \textit{state} contributions, which arise when artificially fixing the matrix elements and only changing the states, and \textit{full} contributions, which arise when changing both (while keeping the same optimized polarizations).

We compare these contributions quantitatively, allowing us to identify genuine regions of parameter space where we can expect to reconstruct the QFI from an optimal ARPES measurement.

\subsection{Materials and responses considered}
\begin{figure}[t!]
    \centering
    \def\svgwidth{0.95\linewidth}
    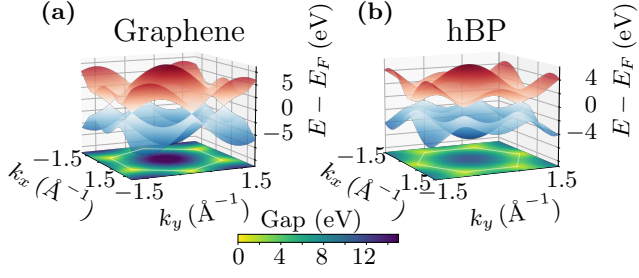
    \caption{Unstrained band and gap structure for the two 2D materials considered. \textbf{(a)} Graphene a gapless semimetal and \textbf{(b)} hexagonal boron phosphide (hBP), a gapped insulator.}
    \label{fig:band_structure}
\end{figure}

We apply our results to two common 2D materials, both of which are well-described by two-band models: graphene and hexagonal boron phosphide (hBP). Both of these have existing tight-binding strain parameterizations \cite{pereira_tight-binding_2009, mortezaei_nobahari_electro-optical_2023}, allowing us to examine the possibility of measuring both the quantum metric $g$ (by setting $\lambda = k_x,k_y)$ and the strain response. For both materials, we evaluate the metric trace Tr$[g] = g_{xx}+g_{yy}$ and the strain response at zero strain $\epsilon$, focusing on the response at $\hbar \omega = 40$\,eV. For the metric trace, we optimize the polarizations for $g_{xx}$ and $g_{yy}$ independently. The band and gap structures at zero strain for both models are shown in Fig.~\ref{fig:band_structure}. Graphene is a semimetal, with gap closings at K and K$'$, whereas hBP is a gapped insulator. Further details on the modelling parameters can be found in App.~\ref{ap:modelling_details_methodology}. Due to availability of data, we use slightly different radial integrals and phase shifts for hBP, but we nonetheless expect all quantitative conclusions to hold.
\section{Results}\label{sec:results}
We now show the results of estimating CFI$_{\hmm}$ from ARPES using the procedure described above, for both the trace of the quantum metric $(\lambda = k_x, k_y)$ and the strain response (setting $\lambda$ to strain). In all of our results, we optimize the polarization vectors $\boldsymbol{e}_1/\boldsymbol{e}_2$ only at the central $\lambda$-value, ensuring that $\hmm$ is independent of $\lambda$. We find, using the optimization procedure described in App.~\ref{ap:finding_optimal_polarizations}, that we are able to construct polarizations that satisfy the equal-magnitude and opposite-axis constraints discussed in Sec.~\ref{sec:photoemission_as_Pauli} at every $\boldsymbol{k}$-point, except at the center of the Brillouin zone ($\Gamma$), where the matrix elements $\boldsymbol{M}_A$ and $\boldsymbol{M}_B$ are not linearly independent. We denote by Tr$[g]_{\mathrm{State}}$ and $\epsilon_{\mathrm{State}}$ the metric trace and strain response obtained by varying only the ground state with $\lambda$ while keeping ARPES matrix elements in Eq.~\eqref{eq:matrix_element_full_expression} at the central $\lambda$, and by Tr$[g]_{\mathrm{Full}}$ and $\epsilon_{\mathrm{Full}}$ the corresponding quantities obtained by varying also the matrix elements.

\subsection{State-only results}
\begin{figure}[t!]
    \centering
    \def\svgwidth{0.95\linewidth}
    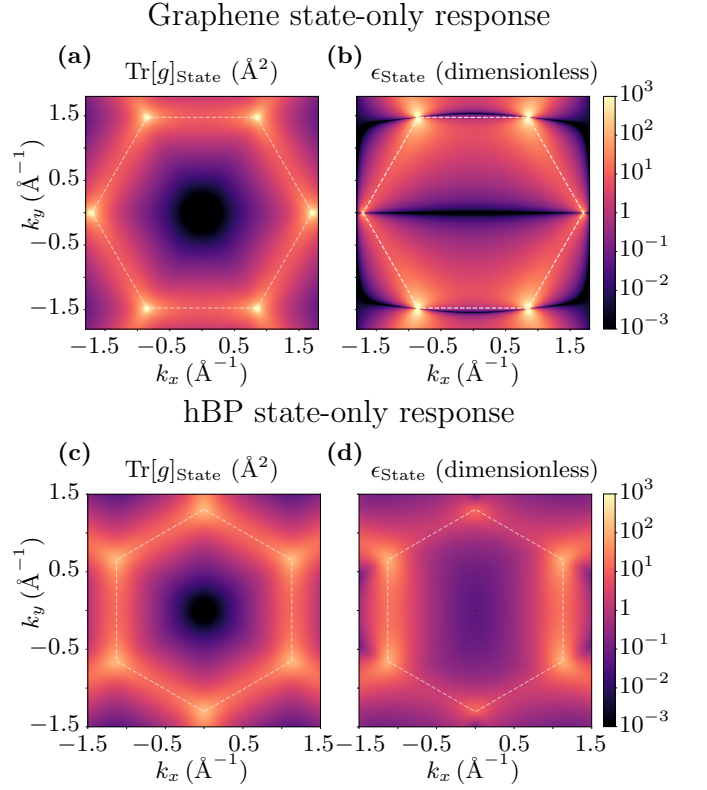
    \caption{ARPES estimated CFI$_{\hmm}$ metric [\textbf{(a)}, \textbf{(c)}] and strain response [\textbf{(b)},\textbf{(d)}] in graphene and hBP, ignoring the parameter dependence of the matrix elements. The results agree perfectly (to within $10^{-7}$) with the optimal QFI from Eq.~\eqref{eq:QFI_def_two_band}, except at $\Gamma$. The optimized incidence and polarization angles are shown in App.~\ref{ap:optimal_angles}.}
    \label{fig:graphene_hbp_fig_1}
\end{figure}

We show the result when optimizing polarizations while keep the matrix elements fixed in Fig.~\ref{fig:graphene_hbp_fig_1}. We find that CFI$_{\hmm}(\boldsymbol{k})$ = QFI$(\boldsymbol{k})$ for both the quantum metric Tr$[g]_{\mathrm{state}}$ and the strain response $\epsilon_{\mathrm{state}}$ for both materials at every $\boldsymbol{k}$-point, except at $\Gamma$. The corresponding optimized incidence and polarization angles are shown in App.~\ref{ap:optimal_angles}.

\subsection{Full results}
\begin{figure*}[t!]
    \centering
    \def\svgwidth{\linewidth}
    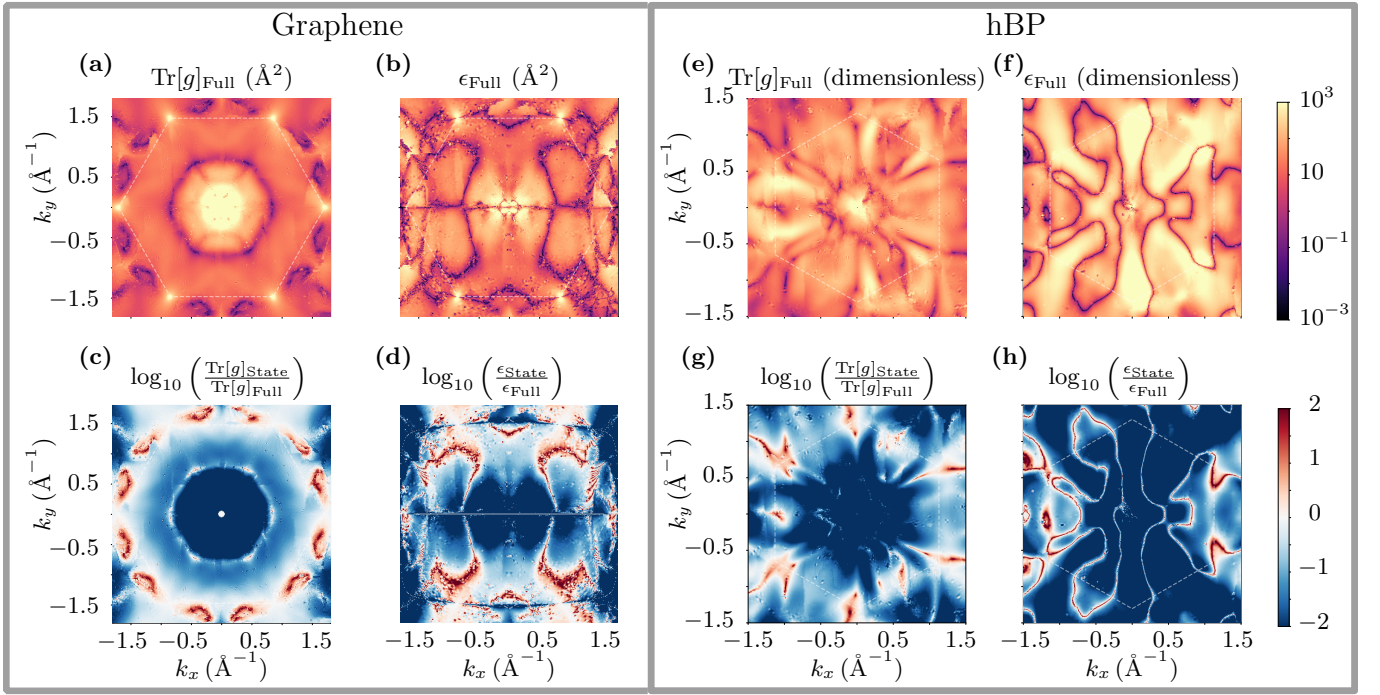
    \caption{ARPES estimated CFI$_{\hmm}$ metric- and strain-responses including variations in the state and matrix elements for graphene \textbf{(a)}--\textbf{(b)} and hBP \textbf{(e)}--\textbf{(f)}. We also show the ratio of the response arising from changes in the quantum state, compared to changes due to the matrix elements for graphene in \textbf{(c)}--\textbf{(d)} and hBP in \textbf{(g)}--\textbf{(h)}.}
    \label{fig:graphene_hbp_full}
\end{figure*}

We show the full metric and strain responses, Tr$[g]_{\mathrm{Full}}$ and $\epsilon_{\mathrm{Full}}$, including variations in the matrix elements for graphene in Fig.~\ref{fig:graphene_hbp_full}(a)--(b) and for hBP in \ref{fig:graphene_hbp_full}(e)--(f). We use the same optimized polarizations at the central $\lambda$-value as in Fig.~\ref{fig:graphene_hbp_fig_1}, ensuring that $\hmm$ is independent of $\lambda$ and that the conditions in Sec.~\ref{sec:photoemission_as_Pauli} are satisfied at all $\boldsymbol{k}$-points for the central $\lambda$-value. We also show the ratios $\mathrm{log}_{10}(\mathrm{Tr}[g]_{\mathrm{State}}/\mathrm{Tr}[g]_{\mathrm{Full}})$ and $\mathrm{\log}_{10}(\epsilon_{\mathrm{State}}/\epsilon_{\mathrm{Full}})$ for graphene in Fig.~\ref{fig:graphene_hbp_full}(c)--(d) and for hBP in \ref{fig:graphene_hbp_full}(g)--(h) respectively. When this ratio is close to zero, QFI $\approx$ CFI$_{\hmm}$ when including all variations, indicating points in $\boldsymbol{k}$-space where we can expect to reliably reconstruct the QFI from ARPES using Eq.~\eqref{eq:CFI_to_D}. 

\section{Discussion}\label{sec:discussion}

\subsection{State-only results}
The results when keeping the ARPES matrix elements fixed at the central $\lambda$-value are shown in Fig.~\ref{fig:graphene_hbp_fig_1}, and agree up to numerical precision with the optimal QFI. We note that for monoatomic materials such as graphene, the wavefunction has only a single independent component at every $\boldsymbol{k}$-point (the sublattice phase). Consequently, the QFI can in principle be estimated from state tomography \cite{kim_direct_2025} by combining multiple dichroism coefficients. This is much more difficult for diatomic materials such as hBP. For our method, there is no difference in complexity between monoatomic and diatomic materials, though we note that for a monoatomic material such as graphene, there exist many different optimal axis as discussed in App.~\ref{ap:optimal_axis_general}, making the space of available polarizations satisfying CFI$_{\hmm}=\mathrm{QFI}$ much larger in such materials.

The optimal angles (shown in App.~\ref{ap:optimal_angles}) vary quite significantly with $\boldsymbol{k}$, indicating that careful control of polarization and incidence may be required to implement this. This arises from the fact that certain transitions are forbidden under certain polarizations and angles. In fact this phenomenon, known as dark corridors in graphene \cite{gierz_illuminating_2011}, can be used to calibrate beam polarization. We note, however, that neither the optimal axis nor the polarization vectors required are generically unique. As such, our optimal angles reflect the results of an inherently underdetermined optimization problem. While attempts have been made to select polarization vectors that give smooth maps in $\boldsymbol{k}$, we only iterated over a small number of possible polarization vectors, and prioritized small incidence angle $\theta_{\mathrm{inc}}$ over smooth angle maps. Furthermore, choosing to fit a different axis than $\partial\hn$ (which may also be optimal or close to optimal) could also improve the angle maps. Finally, we note that the absolute intensity $a$ drops out of our calculations entirely in the ratio in Eq.~\eqref{eq:CFI_to_D}. In practice, however, this value is important as it determines the required integration time. Our procedure in App.~\ref{ap:finding_optimal_polarizations}
optimizes for maximal $a$ when measuring along $\partial \hmm$, but this still results in some points with relatively small $a$ (compared to $||\boldsymbol{M}_A||+||\boldsymbol{M}_B||$). As such, when integration times are a concern, a different axis may be optimal. In experiments, these different constraints would have to be weighted against each other, but we have nonetheless shown that there exist optimal polarizations in each case.
 
\subsection{Full results}
The results when varying both the orbital matrix elements and the eigenvectors with $\lambda$ are shown in Fig.~\ref{fig:graphene_hbp_full}. These results appear significantly more chaotic. This is due to the optimized angles, shown in App.~\ref{ap:optimal_angles}, only being optimized at the central $\lambda$ value. Away from this $\lambda$, $\boldsymbol{M}_A$ and $\boldsymbol{M}_B$ shift, and the dichroism coefficient $D$ can therefore no longer be interpreted as a measurement along $\hmm$. This also explains why the bound $\mathrm{CFI}_{\hmm}\leq \mathrm{QFI}$ is violated in this case. 

Furthermore, we note that the full results in Fig.~\ref{fig:graphene_hbp_full} are not consistently larger than the state-only results in Fig.~\ref{fig:graphene_hbp_fig_1}. This is because of interference effects. While it is possible to look at regions of $\lambda$ where the variations in the state coefficients $C_n^-(\boldsymbol{k})$ are much larger than the variations in the orbital matrix elements (as may be expected e.g. close to a gap-closing point as a function of $\lambda$), this alone is insufficient to guarantee that the measured dichroism gives an estimate of the QFI, as the matrix elements get combined to an overall intensity in Eq.~\eqref{eq:def_ARPES_intensity}, meaning that interference terms can be important even when matrix element variations are small.

Nonetheless, we remark that there for both materials and both responses exist points in the Brillouin-zone where our dichroism based estimate correctly captures the QFI. For the metric Tr$[g]$, this mostly happens where the gap is small, as the variations there dominate. This is less clear in hBP than in graphene, likely because the gap is larger in hBP. The strain responses are less clear, though we note that there also exist regions in the BZ where the full strain response is close to the state-ponly strain response. 

We emphasize that we have only optimized at the central strain value. As such, the regions where $\mathrm{Tr}[g]_{\mathrm{State}} \approx \mathrm{Tr}[g]_{\mathrm{Full}}$ (or equivalently $\epsilon_{\mathrm{State}} \approx \epsilon_{\mathrm{Full}}$), are not a pure material property, but also depend on the chosen polarizations. One could in principle imagine re-optimizing polarizations at every point in the stencil so that it still corresponds to the same $\hmm$ measurement, but obtained from $\{\boldsymbol{M}_A, \boldsymbol{M}_B\}$ at the new stencil point This would require exquisite control of polarization, but should in principle (using our algorithm in App.~\ref{ap:finding_optimal_polarizations}) lead to the full equivalence between QFI and CFI at many points in the BZ. We leave this as an interesting future investigation.

\section{Conclusion and Outlook}\label{sec:Conclusion}
In this work, we have systematically investigated how to measure the Quantum Fisher Information in real materials using Angle-Resolved Photoemission Spectroscopy. We identify under which conditions the ARPES intensity can be related to a Pauli measurement, and how to go from this measurement to the classical and ultimately the quantum Fisher information by carefully tuning incidence and polarizations angles. A fundamental issue with this methodology is that the parameter dependence of the quantum state, captured by the QFI, is difficult to separate from the parameter dependence of the matrix elements. We have investigated this issue quantitatively for strain and momentum derivatives, and have identified regions where one can expect to reconstruc the QFI with this method.

The natural next step is to attempt this in a real experimental setting. This requires good polarization control, but also a good understanding of the underlying photoemission matrix elements, to find the polarizations which result in Pauli measurements. Additionally, we have focused on all matrix elements at a single photon energy, and ignoring spectral functions and energy-conserving $\delta$-functions. Investigating the role of these effects is a extension to our results. Finally, we note the aforementioned idea that it may be possible to independently optimize incidence and polarization angles at every stencil point, which could enable access to the QFI at more $\boldsymbol{k}$-point. The complexity of our scheme compared to other measurement schemes for the QFI remains an interesting future pursuit.

\section*{Acknowledgments}
I thank Javier García de Abajo for access to the \texttt{EDAC} code and Lukasz Plucinski for helpful discussion on APRES and the code. I further thank Jonas Jørgensen Telle and Alv Johan Skarpeid for helpful discussions. The simulations were performed on resources provided by Sigma2 -- the National Infrastructure for High-Performance Computing and Data Storage in Norway. G. F. L. acknowledges funding from the European Union’s Horizon Europe research and innovation programme under the Marie Skłodowska-Curie Grant Agreement No. 101126636, and the YoungCAS fellowship ENQUIRY, awarded by the Centre for Advanced Studies at the Norwegian Academy of Science and Letters.
\bibliography{references_Z}
\clearpage
\newpage
\begin{appendix}

\section{Methodology}\label{ap:modelling_details_methodology}
In this appendix, we discuss details on our calculations leading to Fig.~\ref{fig:graphene_hbp_fig_1} and \ref{fig:graphene_hbp_full} in the main text.
\subsection{EDAC details}
The \texttt{EDAC} code \cite{garcia_de_abajo_multiple_2001} can account for elastic and inelastic multiple scattering of the photoelectron. This is done by considering a cluster of atoms around the emission site, and explicitly considering the scattered wave originated from the emitter site as the final state, including an inelastic mean-free path if desired. We use this code to generate the orbital matrix elements in Eq.~\eqref{eq:atomic_matrix_elements_spherical}, including scattering processes and sublattice phase shifts. 

As we are primarily concerned with illustrating a method, we make some simplifying assumptions in describing the materials.

For graphene, we closely follow \cite{boban_scattering_2025}, with photons at $h\omega  = 40$ eV. We use an inner potential of $V_0 = 17$\,eV, with an inealstic mean free path of $5$\,Å. We find that a scattering order of $30$ produces well-converged plots, without requiring excessive computation, using the recursion method, including up to $l_{\mathrm{max}} = 15$ components. We use an angular grid resolution of $1^{\circ}$, and interpolate the matrix elements to a rectangular grid. We do the computation for a cluster going up to $R = 20$\,Å from the central unit site, resulting in a total of $482$ atoms in the cluster at zero strain. The radial wavefunctions and phase shifts were evaluted using tabulated values for graphene $2p$ orbitals from \cite{goldberg_photoionization_1981}. We used a work function $W = 6$ eV.

\begin{figure}[t!]
    \centering
    \def\svgwidth{\linewidth}
    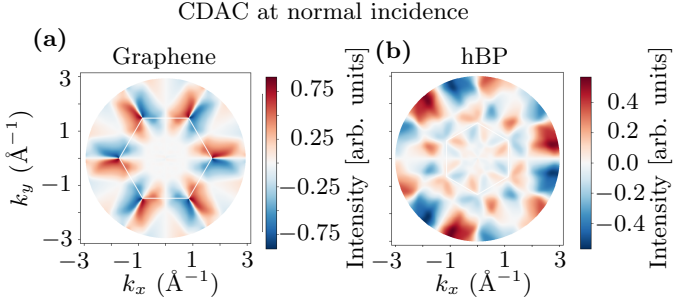
    \caption{The computed circular dichroism coefficient  CDAC$=(I_{\mathrm{LCP}}-I_{\mathrm{RCP}})/(I_{\mathrm{LCP}}+I_{\mathrm{RCP}})$ for \textbf{(a)} grapene and \textbf{(b)} hBP at normal incidence.}
    \label{fig:CDAC_app}
\end{figure}

For hBP, we use the same setup, but with an inner potential of 10\,eV \cite{ohtake_atomic_2023}. Also, as the radial wavefunctions and phase shifts for B and P $2p$ orbitals are not available from \cite{goldberg_photoionization_1981}, we instead use those from respectively Ga $4p$ and As $4p$ instead, which can therefore be thought of as slightly heavier versions of the real material. Going up to $R= 20$\,Å gives a total of $422$ atoms in this case. As an example, we show the circular dichroism coefficient at normal incidence in Fig.~\ref{fig:CDAC_app}, with the graphene case agreeing with \cite{boban_scattering_2025}.

\subsection{Model details}
For the graphene strain response, we implement the model from \cite{pereira_tight-binding_2009}. The strain response is evaluated at zero strain and $\theta = 0$. For hBP, we use the strain parameterization from \cite{mortezaei_nobahari_electro-optical_2023}, and consider the response with respect to $\epsilon_{xx}$. For strain responses, we recompute the orbital photoemission matrix elements at finite strain by changing the shape of the cluster, but do not change the inner potential or scattering lengths. This is justified as we are considering the response close to zero strain. 

\subsection{Orbital embedding dependence of the procedure}
A subtlety related to this formulation of ARPES matrix elements is the choice of orbital embedding \cite{telle_optimally_2026}. Our formulation assumes a common basis choice between the tight-binding model and the matrix elements, as otherwise the vectors $\hn$ and $\hmm$ are not well-defined. As we are using an orbital-embedded basis in Eq.~\eqref{eq:atomic_matrix_elements_spherical}, straining the system changes the basis. To get a consistent comparison in the case where we fix the matrix elements but vary the strain, we therefore adjust the embedding.

\subsection{Finding the site-resolved photoemission matrix elements}\label{app:edac-mu-conversion}
Finding the optimal polarizations requires us to evaluate the matrix elements $M_{in} = \bra{\psi_f}\hat{r}_i\ket{\phi_n}$ for the orbitals in the unit cell. The \texttt{EDAC} code gives access to the (real and complex part of) these matrix elements for any given polarizations. To reconstruct the general matrix elements (in a spherical basis), we evaluate \texttt{EDAC} with three linearly independent polarizations for each orbital in the unit cell.  The angle and polarization convention used by the code are: \begin{equation}\label{eq:polarization_vector_parameterization_mu}
 \boldsymbol\epsilon_\mu = (\epsilon_{+},\epsilon_0,\epsilon_{-}),
\end{equation}
with
\begin{align}\label{eq:definition_polarization_vectors}
\begin{split}
 \epsilon_{+} &= \frac{\left(\cos\alpha\cos\theta_{\rm inc} + i\sin\alpha\,e^{i\delta}\right)e^{i\phi_{\rm inc}}}{\sqrt{2}},\\
 \epsilon_{0} &= -\cos\alpha\sin\theta_{\rm inc},\\
 \epsilon_{-} &= \frac{\left(-\cos\alpha\cos\theta_{\rm inc} + i\sin\alpha\,e^{i\delta}\right)e^{-i\phi_{\rm inc}}}{\sqrt{2}},
\end{split}
\end{align}
where $(\theta_{\mathrm{inc}}, \phi_{\mathrm{inc}})$ specify the plane of the incoming photon and $(\alpha,\delta)$ specify the polarization properties of the incoming photon. We choose the three linearly independent polarization vectors to be:
\begin{align*}
\begin{split}
 \boldsymbol\epsilon^{(\mathrm{LP1})} &= \boldsymbol\epsilon_\mu\bigl(\theta_{\rm inc}=\pi/3,\phi_{\rm inc}=0,\alpha=0,\delta=0\bigr),\\
 \boldsymbol\epsilon^{(\mathrm{LP2})} &= \boldsymbol\epsilon_\mu\bigl(\theta_{\rm inc}=\pi/3,\phi_{\rm inc}=\pi/2,\alpha=0,\delta=0\bigr),\\
 \boldsymbol\epsilon^{(\mathrm{LCP})} &= \boldsymbol\epsilon_\mu\bigl(\theta_{\rm inc}=\pi/3,\phi_{\rm inc}=0,\alpha=\pi/4,\delta=\pi/2\bigr).
\end{split}
\end{align*}
Which we collect into the polarization matrix $\boldsymbol{E}=[\boldsymbol\epsilon^{(\mathrm{LP1})},\boldsymbol\epsilon^{(\mathrm{LP2})},\boldsymbol\epsilon^{(\mathrm{LCP})}]^T$. At each $\boldsymbol{k}$-point and for each atom in the uit cell, we collect the three measured amplitudes into a vector $\mathbf{m}_{\rm meas}^n =[
  m_{\rm LP1}^n,
  m_{\rm LP2}^n,
  m_{\rm LCP}^n]^T$. These are related to the orbital matrix elements $\boldsymbol{M}_n$ by:
\begin{equation}
 \mathbf{m}_{\rm meas}^n = \boldsymbol{E}\boldsymbol{M}_n
 \qquad
 \boldsymbol{M}_n =
 \begin{pmatrix}
  M_{+}^n\\ M_0^n\\ M_{-}^n
 \end{pmatrix}.
\end{equation}
We finally perform the reconstruction pointwise by
\begin{equation}
 \boldsymbol{M}_n(\boldsymbol{k}) = \boldsymbol{E}^{-1}\,\mathbf{m}_{\rm meas}^n(\boldsymbol{k}).
\end{equation}

This gives us access to the orbital matrix elements in the spherical basis at every $\boldsymbol{k}$ point. A final technical detail is that \texttt{EDAC} internally flips the sign of $\alpha$ provided by the user, which has to be taken into account. 

\section{Finding the optimal polarizations}\label{ap:finding_optimal_polarizations}
We now address the central optimization problem: how to find two complex polarizations vectors $\boldsymbol{e}_{1,2}$ that satisfy the equal magnitude and opposite axis constraints, discussed in Sec.~\ref{sec:photoemission_as_Pauli}, which make it possible to interpret the dichroism coefficient as a Pauli measurement. Before going into the details, we note that there are two additional requirements: first of all, we want the strength of the electromagnetic field to be constant, so we require $|\boldsymbol{e}_1|^2 = |\boldsymbol{e}_2|^2 = 1$. Furthermore, we want the resultant polarization vector to be representable in the form given
in Eq.~\ref{eq:definition_polarization_vectors}, and we want a small incidence angle $\theta_{\mathrm{inc}}$ to ensure large intensities.

Here we describe how we can find polarization vectors $\boldsymbol{e}_{1,2}$ that fullfill the target conditions of (a) balanced intensities $a_1 = a_2$ and (b) opposite axis $\hmm_1 = -\hmm_2$ (or equivalently $\boldsymbol{g}_1 = -\boldsymbol{g}_2$) and (c) normalized polarizations $|\boldsymbol{e}_1|^2 = |\boldsymbol{e}_2|^2$ efficiently as long as $\boldsymbol{M}_A\nparallel \boldsymbol{M}_B$. Mapping these back to experimentally realizable polarizations will be discussed in the subsequent section.

\subsection{Finding optimal polarization vectors}\label{ap:optimal_polarization_vectors}
We start by noting that the matrix elements $\boldsymbol{M}_A, \boldsymbol{M}_B$ with $M_A^i = \boldsymbol{e}_i\cdot\boldsymbol{M}_A$ and $M_B^i = \boldsymbol{e}_i\cdot\boldsymbol{M}_B$ are three-dimensional vectors with complex coefficients that span a two-dimensional complex subspace of the three-dimensional polarization space. While they are not gauge invariant, any change in gauge only results in an overall phase for each vector. We employ the Gram-Schmidt procedure to construct a Hermitian orthonormal basis ($\boldsymbol{\alpha}_1,\boldsymbol{\alpha}_2$) for this subspace. Crucially for our construction, however, polarization vectors not in $\mathrm{span}_{\mathbb{C}}\{\boldsymbol{\alpha}_1, \boldsymbol{\alpha}_2\}$ can still contribute to the final matrix elements, as the matrix elements are constructed from the bilinear product $M_n^i=\boldsymbol{e}_i^T\boldsymbol{M}_n$, rather than the hermitian product $\boldsymbol{e_i}^{\dagger}\boldsymbol{M}_{n}$. We can therefore also consider a third normalized complex vector $\boldsymbol{\alpha}_3$ in the orthogonal complement of the span, completing the span of polarization space, which can contribute to the matrix element. We will use this third vector to ensure normalization. Defining $\boldsymbol{\alpha}=(\boldsymbol{\alpha}_1,\boldsymbol{\alpha}_2, \boldsymbol{\alpha}_3)$, any desired polarization vector $\boldsymbol{e}_i$ can be written as $\boldsymbol{e}_i = \boldsymbol{c}_i^T\boldsymbol{\alpha}$, for some coefficients $\boldsymbol{c}$. Note that $||\boldsymbol{e}_i|| = ||\boldsymbol{c}_i||$.

The (unormalized) effects in Eq.~\eqref{eq:effects} for a given polarization $\boldsymbol{v}_i = (M_A^i, M_B^i)^T$ are now given by $\boldsymbol{v}_i=\boldsymbol{L}\boldsymbol{c}_i$ with:
\begin{equation}\label{eq:optimization_matrix_2x3}
    \boldsymbol{L}= \begin{pmatrix}
        \boldsymbol{\alpha}_1^T\boldsymbol{M}_A & \boldsymbol{\alpha_2}^T\boldsymbol{M}_A & \boldsymbol{\alpha_3}^T\boldsymbol{M}_A\\
        \boldsymbol{\alpha}_1^T\boldsymbol{M}_B & \boldsymbol{\alpha_2}^T\boldsymbol{M}_B & \boldsymbol{\alpha}_3^T\boldsymbol{M}_B
    \end{pmatrix}.
\end{equation}
The desired $\boldsymbol{v}_i$'s are determined by the desired axis. In practice, we align $\boldsymbol{v}$ to be along $\partial_{\lambda}\hn$. We can then invert $\boldsymbol{L}$ to find the coefficients $\boldsymbol{c}_i$ which realize the required $\boldsymbol{v}_i$. As $\boldsymbol{v}_i$ has only two components, but there are $3$ independent components of $\boldsymbol{c}_i$, the extra degree of freedom can be used to fix the normalization.

Specifically, a given desired axis $g$ results in a given normalized two-component complex vector $\boldsymbol{u}(g)$ from Eq.~\eqref{eq:effects}, which can be found through stereographic projection, i.e. writing $\boldsymbol{g}=(\sin\tilde{\theta}\cos \tilde{\varphi}, \sin \tilde{\theta} \sin \tilde{\varphi}, \cos \tilde{\theta})$ gives (up to a global phase):
\begin{equation}
\boldsymbol{u}(g) = \begin{pmatrix}
    \cos(\tilde{\theta}/2)e^{i\tilde{\varphi}} \\
    \sin(\tilde{\theta}/2)
\end{pmatrix}, \boldsymbol{u}(-g) = \begin{pmatrix}
-\sin(\tilde{\theta}/2)e^{i\tilde{\varphi}}\\
\cos(\tilde{\theta}/2)
\end{pmatrix},
\end{equation}
We then seek two effects $\boldsymbol{v}_1 = \sqrt{2a_0}\boldsymbol{u}(g)$ and $\boldsymbol{v}_2 = \sqrt{2a_0}\boldsymbol{u}(-g)$, which are guaranteed to satisfy the equal magnitude and opposite axis constraint. We note that this means we can in pracrice target any axis.

Knowing the target effects, we can invert Eq.~\ref{eq:optimization_matrix_2x3}, as $\boldsymbol{c}_i = \boldsymbol{L}^{+}\boldsymbol{v}_i$, where $\boldsymbol{L}^+$ is the pseudoinverse of $\boldsymbol{L}$. As we want the polarization to be normalized, we now ask what the largest $a_0$ (i.e. maximal intensity) that we can get is, while retaining normalized polarizations $||\boldsymbol{c}||=1$.  Decomposing $\boldsymbol{c}_i = \boldsymbol{c}_i^{\parallel}+\lambda_i\boldsymbol{c}_i^{\perp}$ with $\lambda_i\in \mathbb{C}$, $||\boldsymbol{c}_{\perp}|| = 1$ and $\boldsymbol{L}\boldsymbol{c}_i^{\perp} = \boldsymbol{0}$, gives $||\boldsymbol{c}_i||^2 = ||\boldsymbol{c}_i^{\parallel}||^2+|\lambda_i|^2$, so that the minimal norm is $||\boldsymbol{c}_i^{\parallel}|| = ||\boldsymbol{L}^+\boldsymbol{v}_i|| = \sqrt{2a_0}||\boldsymbol{L^+}\boldsymbol{u}||$. A solution therefore exists as long as $||\boldsymbol{c}_i^{\parallel}|| \leq 1$, giving the condition:
\begin{equation}
    a_{0, \mathrm{max}}^i = \frac{1}{2||\boldsymbol{L}^+\boldsymbol{u}_i||}.
\end{equation}
And choosing $a_0$ to be $a_0 = \mathrm{min}\{ a_{0, \mathrm{max}}^1,a_{0, \mathrm{max}}^2\}$, ensures that both polarization vectors can be physically reproduced with unit norm. 

There is now a whole family of solutions which satisfy the equal magnitude and opposite axis constraint while being normalized, which can be parameterized by:
\begin{equation}
    \boldsymbol{c}_i = \boldsymbol{L}^+\boldsymbol{u}_i + \sqrt{1-||\boldsymbol{L}^+\boldsymbol{u}_i||^2}e^{i\phi_i}\boldsymbol{n}.
\end{equation}
For a fixed $\phi_i$, the associated polarization is then finally $\boldsymbol{e}_i = \boldsymbol{c}_i^T\boldsymbol{\alpha}$. There is still a global gauge freedom in this expression, which we fix by ensuring that the $\mu=0$ component is real, to be consistent with Eq.~\eqref{eq:definition_polarization_vectors}.

This shows that there are many polarization vectors which satisfy all constraints. In practice, we iterate over a few $\lambda$-values, and prioritize those polarizations which lead to (a) a small incidence angle $\theta_{\mathrm{inc}}$ for physical realizability and (b) a smooth spatial dependence of the angles, as discussed in the next section.

\subsection{Mapping polarization to experimental angles}\label{app:mapping-to-angles}
Having found the optimal polarization vectors $\boldsymbol{e}_{1,2}$ as described in App.~\ref{ap:optimal_polarization_vectors}, we now ask how to map these back to the experimental incidence angles and polarization vectors.

We saw in Eq.~\eqref{eq:definition_polarization_vectors} how a choice of $(\theta_{\mathrm{inc}}, \phi_{\mathrm{inc}},\alpha,\delta)$ leads to a polarization vector $\boldsymbol{e}$. To invert this relation, we use \texttt{Scipy}'s \cite{virtanen_scipy_2020} least square fitting mwrhos, with a few different starting points to avoid local minima. 

We perform an unconstrained fit and then map the resultant angles back to their principal branch. We then fold the resultant angles into the ranges $\theta_{\mathrm{inc}}\in [0,\pi], \phi_{\mathrm{inc}}\in [-\pi,\pi], \alpha\in [-\pi/2, \pi/2]$ and $\delta \in [-\pi, \pi]$. This can always be done as the polarization vector is invariant under the mappings:
\begin{equation}
    (\theta_{\mathrm{inc}}, \phi_{\mathrm{inc}}, \alpha, \delta)\rightarrow (-\theta_{\mathrm{inc}}, \phi_{\mathrm{inc}}+\pi, \pi-\alpha, \delta+\pi),
\end{equation}
and only flips an overall sign (corresponding to a gauge transformation) under:
\begin{equation}
    (\theta_{\mathrm{inc}}, \phi_{\mathrm{inc}}, \alpha, \delta)\rightarrow (\pi-\theta_{\mathrm{inc}}, \phi_{\mathrm{inc}}+\pi, \pi-\alpha, \delta).
\end{equation}
These transformations can be used to force the angles into the desired range. Numerically, we find that the polarization vector arising from the fitted angles $\boldsymbol{\mu}$ agree with the desired polarization vector $\boldsymbol{e}$ (up to an overall sign $s$) as $||\boldsymbol{\mu}-s\boldsymbol{e}|| < 10^{-6}$ for all materials, physical response and $\boldsymbol{k}$. We show the fitted angles in App.~\ref{ap:optimal_angles}.

\section{Further details on optimal axis}\label{ap:optimal_axis_general}
While we in the main text always chose the axis $\hmm$ (at $\lambda$) to be along $\partial_{\lambda}\hn$, we note that this choice is not unique, and there may exist many possible optimal axes. This is important if one wants to optimize all parameters in the problem for experimental feasibility.

In particular, starting from Eq.~\eqref{eq:CFI_general_axis} in the main text, and assuming that the axis $\hmm$ is not parallel to the Bloch vector $\hn$, i.e.  $(\hmm\cdot \hn)^2\neq 1$, $\hmm$ can be decomposed as $\hmm= (\hmm\cdot \hn)\hn+\hmm_{\perp}$, with $\hmm_{\perp}\cdot\hn = 0$. Defining $(\partial \hn \cdot \hmm_{\perp}) = ||\partial \hn||||\hmm_{\perp}||\cos \theta$ and using that $|\hmm_{\perp}|^2 = 1-(\hn\cdot \hmm)^2$ gives:
\begin{equation}
\mathrm{CFI}_{\hmm}=||\partial \hn||^2\cos^2\theta.
\end{equation}
By definition, $\hn\cdot \hmm_{\perp}= 0$ and $\hn\cdot \partial \hn = 0$. This implies for a fixed measurement axis $\hmm$ that if $\hn, \partial\hn$ and $\hmm$ all lie in the same 2D plane, and $|\hmm_{\perp}| \neq 0 $, then $\partial\hn||\hmm_{\perp}$, and therefore $\cos \theta = 1$. In this situation, CFI$_{\hmm} = \mathrm{QFI}$. 

There is therefore a simple geometric condition for when $\hmm$ is an optimal measurement direction: whenever there exists a locally constant (in $\lambda$) 2D plane spanned by $\{\hn, \partial \hn\}$, then any measurement axis $\hmm$ that lies in this plane and not along $\hn$ is optimal. The most useful example of such a situation is when a symmetry of the system forces a component of the Bloch vector to be zero, and the perturbation $\lambda$ preserves this symmetry. This is the case for graphene, where sublattice symmetry ensures that $n_z = 0$. Consequently, any measurement axis in the $xy-$plane not along $\hn (\boldsymbol{k})$ will constitute an optimal measurement. This also shows that there is great freedom in optimizing $\hmm$ to satisfy experimental constraints.

\section{Optimal angles}\label{ap:optimal_angles}
The optimal angles for all parameters and materials are shown in Fig.~\ref{fig:optimal_angles_strain_graphene}--\ref{fig:optimal_angles_ky_hbp}.
\begin{widetext}
\newpage
\begin{figure*}[ht!]
    \centering
    \def\svgwidth{\linewidth}
    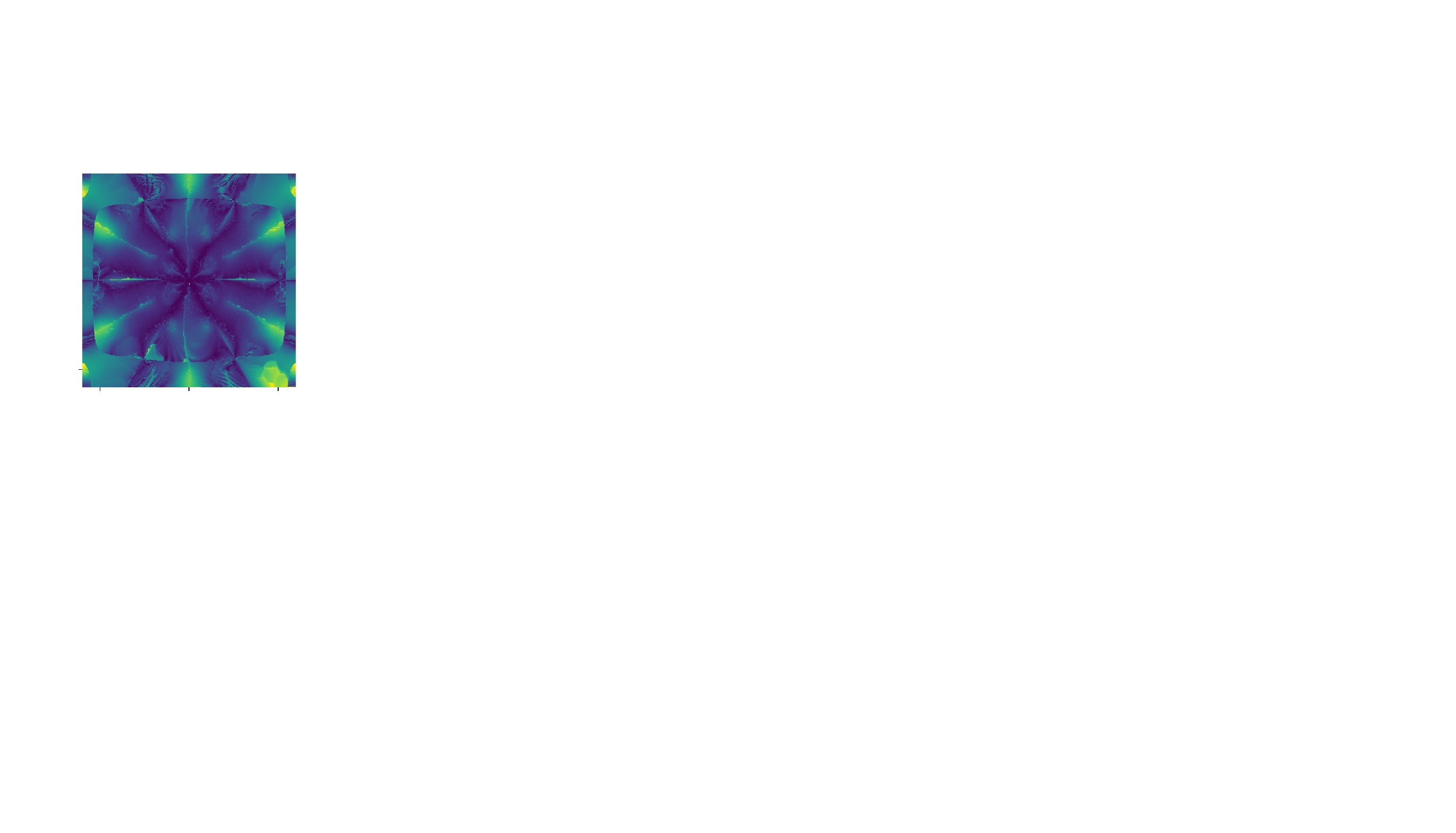
    \caption{Optimal angle (in radians) for the strain response in graphene.}
    \label{fig:optimal_angles_strain_graphene}
\end{figure*}
\begin{figure*}[b!]
    \centering
    \def\svgwidth{\linewidth}
    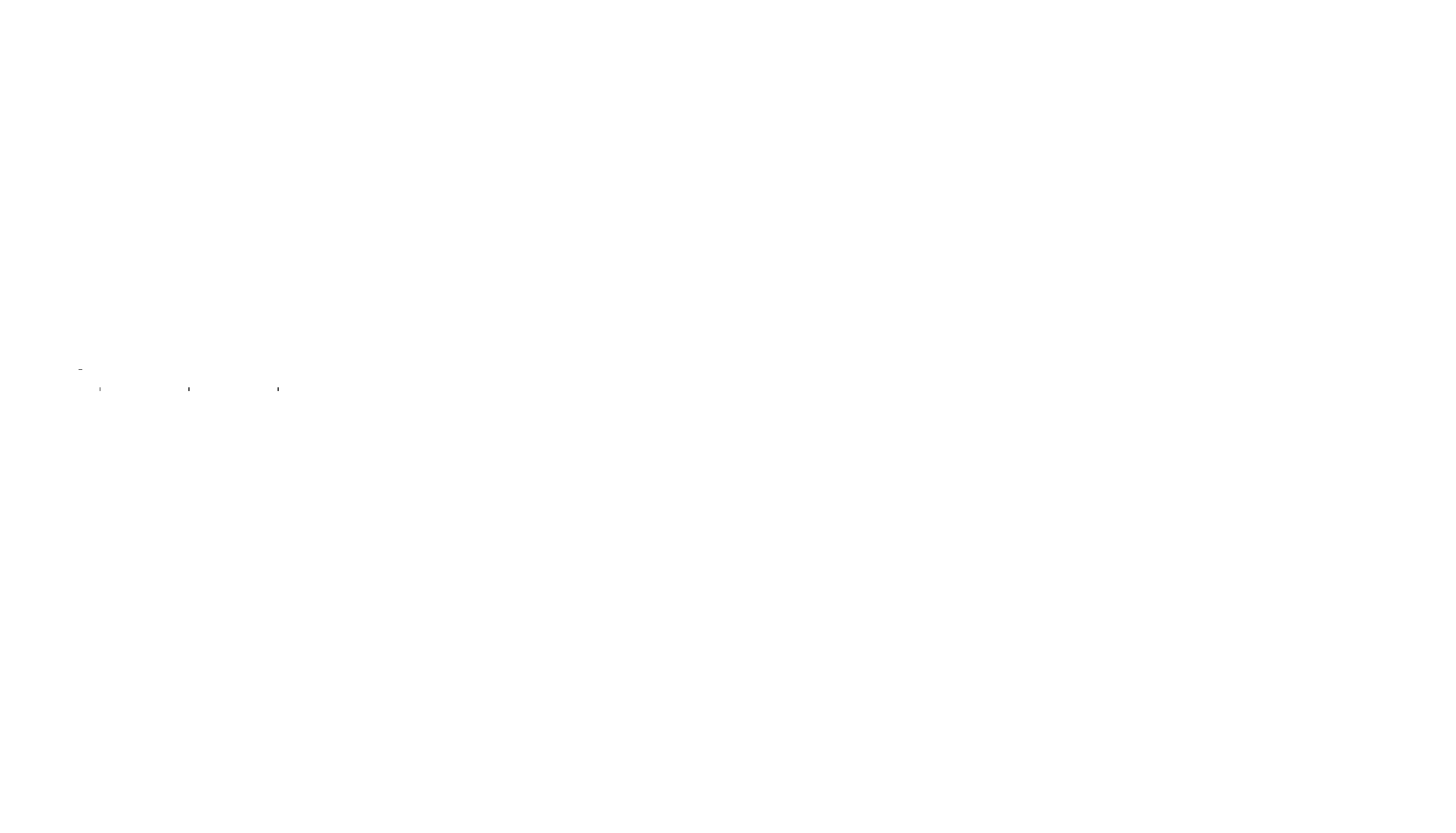
    \caption{Optimal angles (in radians) for $g_{xx}$ in graphene.}
    \label{fig:optimal_angles_kx_graphene}
\end{figure*}
\begin{figure*}[ht!]
    \centering
    \def\svgwidth{\linewidth}
    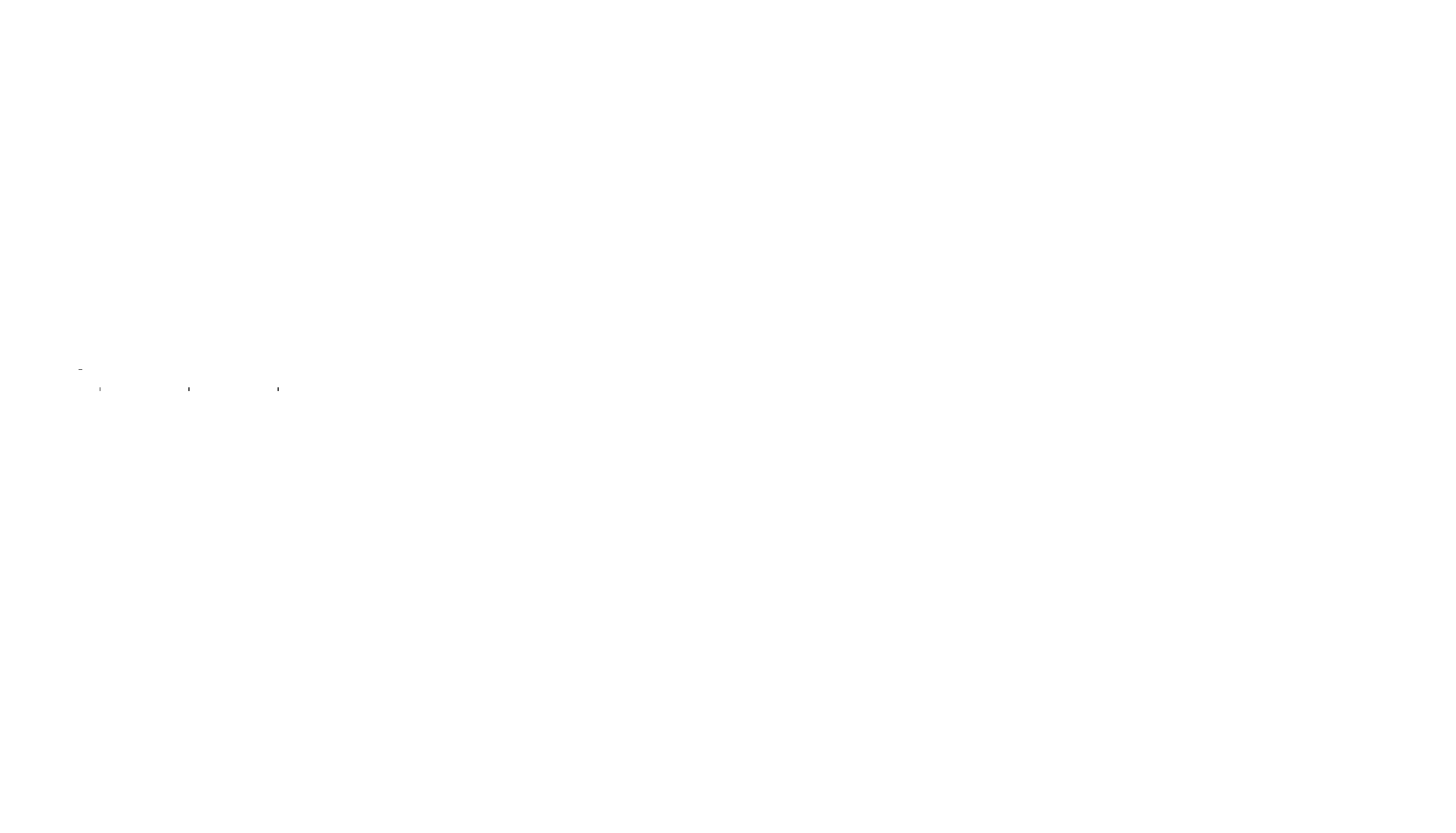
    \caption{Optimal angles (in radians) for $g_{yy}$ in graphene.}
    \label{fig:optimal_angles_ky_graphene}
\end{figure*}

\begin{figure*}[t!]
    \centering
    \def\svgwidth{\linewidth}
    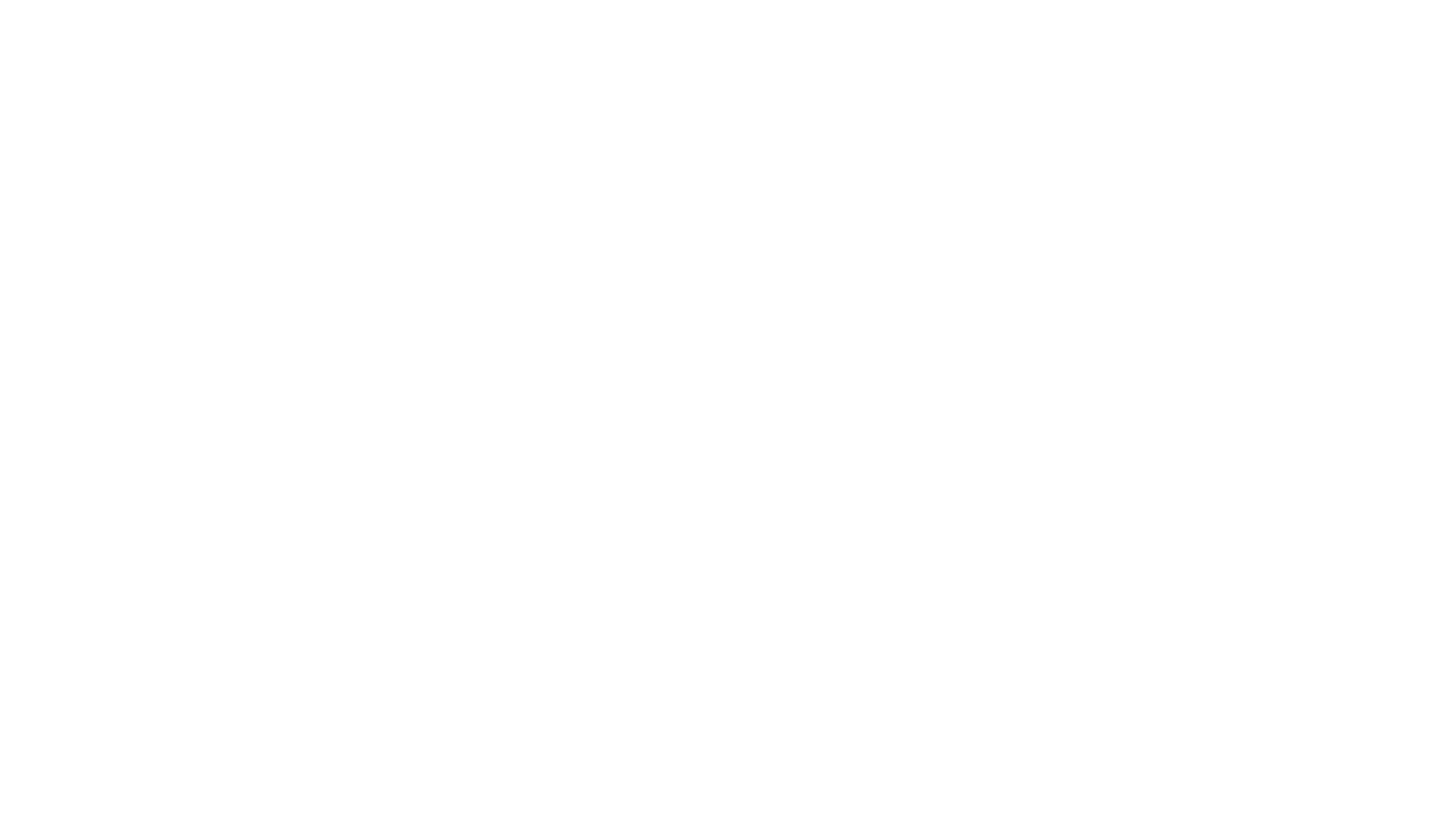
    \caption{Optimal angles (in radians) for strain in hBP.}
    \label{fig:optimal_angles_strain_hbp}
\end{figure*}
\begin{figure*}[t!]
    \centering
    \def\svgwidth{\linewidth}
    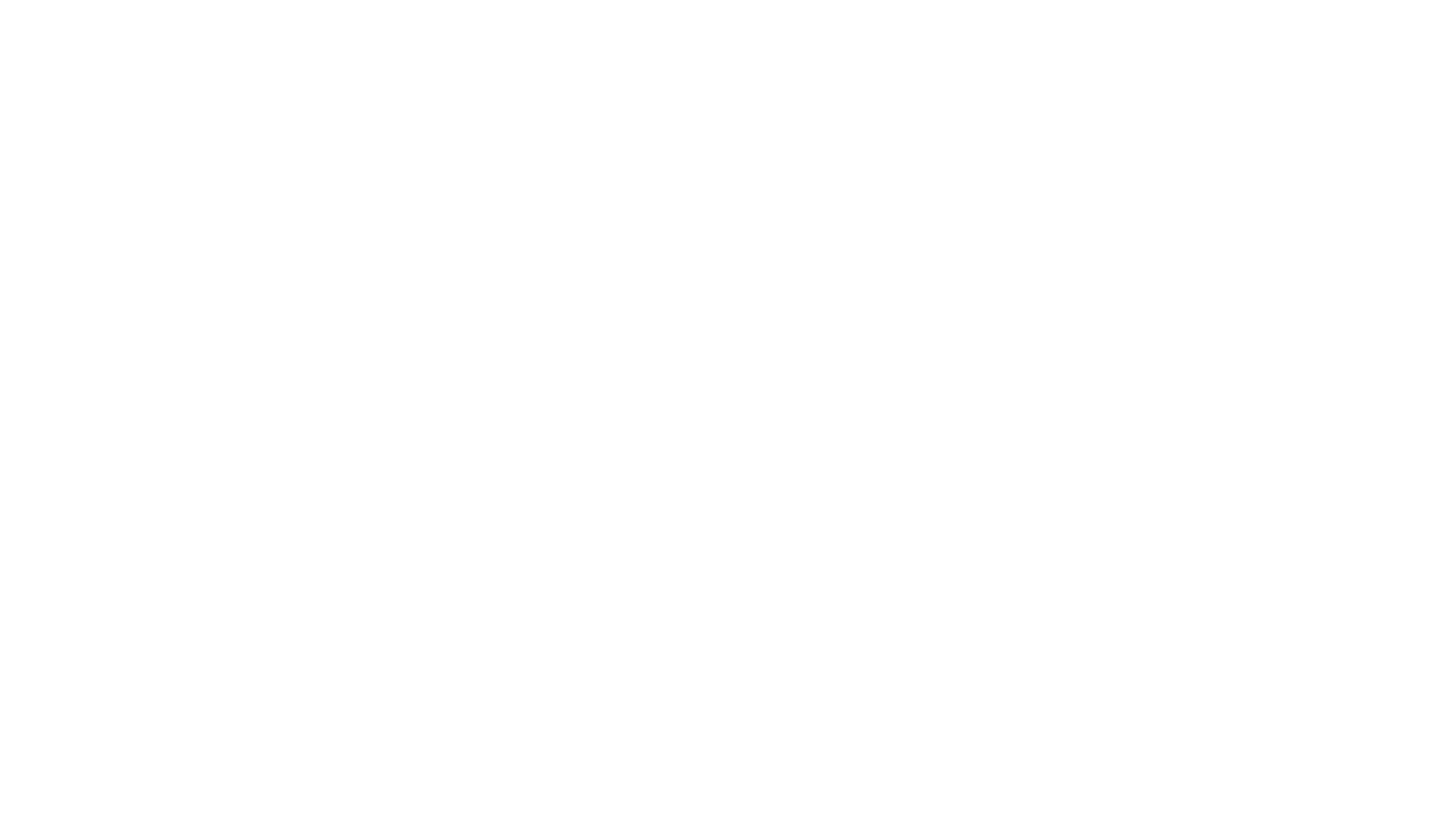
    \caption{Optimal angles (in radians) for $g_{xx}$ in hBP.}
    \label{fig:optimal_angles_kx_hbp}
\end{figure*}
\begin{figure*}[t!]
    \centering
    \def\svgwidth{\linewidth}
    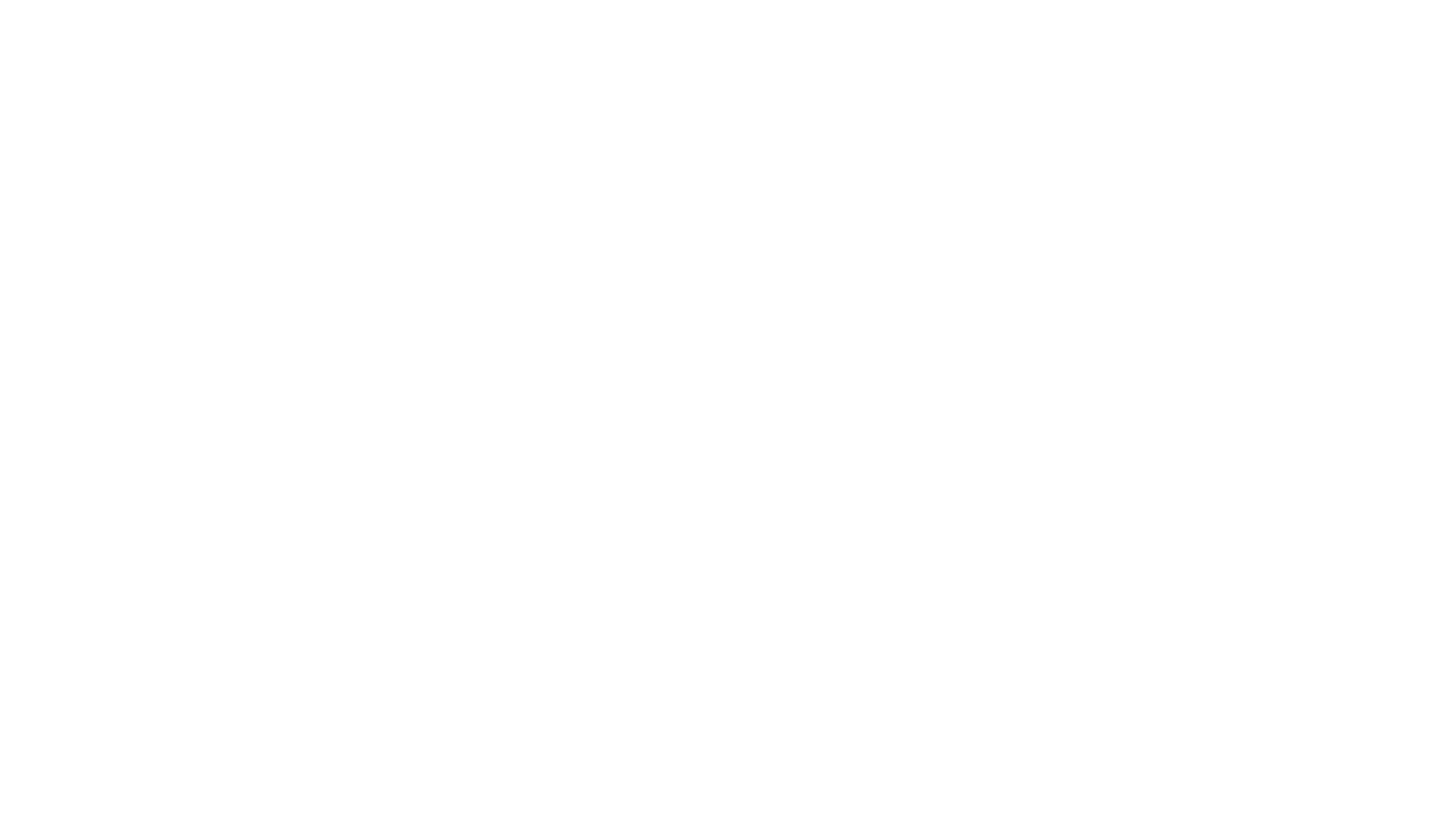
    \caption{Optimal angles (in radians) for $g_{yy}$ in hBP.}
    \label{fig:optimal_angles_ky_hbp}
\end{figure*}
\end{widetext}
\end{appendix}

\end{document}